\documentclass[a4 paper, 12 pt] {article}

\usepackage{graphics, graphicx}

\begin{document}

\title{\bf{Generalized Eikonal Knots and New Integrable Dynamical Systems}}
\author{A. Wereszczy\'{n}ski  \thanks{wereszczynski@th.if.uj.edu.pl}
       \\
       \\ Institute of Physics,  Jagiellonian University,
       \\ Reymonta 4, 30-059 Krak\'{o}w, Poland}
\maketitle
\begin{abstract}
A new class of non-linear $O(3)$ models is introduced. It is shown
that these systems lead to integrable submodels if an additional
integrability condition (so called the generalized eikonal
equation) is imposed. In the case of particular members of the
family of the models the exact solutions describing toroidal
solitons with a non-trivial value of the Hopf index are obtained.
Moreover, the generalized eikonal equation is analyzed in detail.
Topological solutions describing torus knots are presented.
Multi-knot configurations are found as well.
\end{abstract}
\newpage
\section{\bf{Introduction}}
It has been recently shown that the complex eikonal equation
\begin{equation}
\partial_{\mu} u \partial^{\mu} u=0 \label{eikonal}
\end{equation}
plays a prominent role in nonlinear field theories in higher
dimensions. Such models, widely applied in many physical contexts,
are not, in general, integrable. In a consequence the spectrum of
nontrivial (f.ex. topological) solutions is scarcely known.
Fortunately, the eikonal equation allows us to define integrable
subsectors for such models with not empty set of the topological
solutions \cite{alvarez}, \cite{fujii}, \cite{aratyn},
\cite{ferreira}, \cite{sanchez1}, \cite{sanchez2}.
\\
The nonlinear $O(3)$ models in two and three space dimensions can
serve as the best example. These models, original based on an
unit, three component vector field, can be reformulated in terms
of a complex field $u$ by means of the standard stereographic
projection
\begin{equation}
\vec{n}= \frac{1}{1+|u|^2} ( u+u^*, -i(u-u^*), |u|^2-1).
\label{stereograf1}
\end{equation}
As the static, finite energy configurations are nothing else but
maps from the compactified $R^2$ or $R^3$ on $S^2$, they can be
classified by the pertinent topological invariant: the winding
number or the Hopf index $Q_H$ respectively. As a result solutions
describing topological vortices or knots can be obtained. The
first model possessing the simplest knot soliton i.e. hopfion, is
referred as the Nicole model \cite{nicole}. Moreover, it has been
proved that this solution fulfills additionally an integrability
condition i.e. the eikonal equation. The idea that hopfions can be
quite easily found in integrable subsectors has enabled us to
construct a family of the generalized Nicole models where hopfions
which higher topological charges have been reported \cite{ja
nicole}. Strictly speaking all these hopfions are identical to the
eikonal knots \cite{adam}, \cite{ja eikonal} being solutions of
the complex eikonal equation in the toroidal coordinates
\footnote{The eikonal equation admits also other topological
objects f.ex. vortices, braided strings or hedgehogs \cite{ja
other}}
$$ x=\frac{\tilde{a}}{q} \sinh \eta \cos \phi , $$
$$ y=\frac{\tilde{a}}{q} \sinh \eta \sin \phi , $$
\begin{equation}
z=\frac{\tilde{a}}{q} \sin \xi ,\label{tor_coord}
\end{equation}
where $q=\cosh \xi - \cos \phi $ and $\tilde{a}$ is a scale
constant. In fact it has been observed by Adam \cite{adam} that
the following function
\begin{equation}
u(\eta, \xi, \phi) =f_0 (\eta) e^{i(m\xi +k\phi)}, \label{sol_eik}
\end{equation}
where
\begin{equation}
f_0(\eta) = A \sinh^{\pm |k|} \eta \frac{\left( |m| \cosh \eta +
\sqrt{k^2 +m^2\sinh^2 \eta } \right)^{\pm |m|}}{\left( |k| \cosh
\eta + \sqrt{k^2 +m^2\sinh^2 \eta } \right)^{\pm |k|}},
\end{equation}
solves (\ref{eikonal}) and carries non-vanishing topological
charge
\begin{equation}
Q_H= \pm |mk|. \label{charge}
\end{equation}
Here $m,k$ are integer numbers and $A$ is a complex constant.
Using the group of the target symmetries \cite{adam new} one can
construct even more general solution
\begin{equation}
\tilde{u} = F(u), \label{sol eik ogolne}
\end{equation}
where $F$ is an arbitrary holomorphic function \cite{adam},
\cite{ja eikonal}.
\\
It should be emphasized that the eikonal knots are not only
solutions of the toy models but can also find a physical
application as approximated solutions of the Faddeev-Niemi
effective model \cite{niemi} for the low energy gluodynamics
\cite{inni}. They provide an analytical framework in which
qualitative (shape and topology) as well as quantitative (energy)
features of the Faddeev-Niemi hopfions can be captured \cite{ja
eikonal}, \cite{ja FN}.
\\
The fact that the complex eikonal equation has appeared to be very
helpful in deriving hopfions in the Nicole-type models might
indicate that in the case of other non-integrable $O(3)$ models a
similar condition might be introduced.
\\
The main aim of the present paper is to define a generalization of
the eikonal equation which will enable us to derive new integrable
submodels, for which new analytical toroidal solitons might be
found.
\section{\bf{Generalized eikonal knots}}
The original eikonal equation appears in the context of the
nonlinear $O(3)$ model by the following vector quantity
\begin{equation}
K_{\mu}^{(1)}=\partial_{\mu} u. \label{k1}
\end{equation}
In order to guarantee the existence of an integrable submodel this
object must obey two conditions \cite{aratyn}
\begin{equation}
K_{\mu}^{(i)} \partial^{\mu} u =0 \label{cond int1}
\end{equation}
and
\begin{equation}
\mbox{Im} ( K_{\mu}^{(i)} \partial^{\mu} u^*) =0, \label{cond
int2}
\end{equation}
where $i=1,2..$ (see below). It is straightforward to notice that
equation (\ref{cond int1}) gives the eikonal equation. Using
$K^{(1)}_{\mu}$ one can construct a dynamical model (but
non-integrable) i.e. the Nicole model \cite{nicole}
\begin{equation}
L=\frac{1}{(1+|u|^2)^3} (K_{\mu}^{(1)} \partial^{\mu}
u^*)^{\frac{3}{2}}, \label{nicole}
\end{equation}
where the power $3/2$ is chosen to omit the Derick scaling
argument for nonexistence of the topological solitons. Now,
equation (\ref{cond int1}) is an integrability condition for the
model (for details see \cite{nicole}, \cite{ja nicole}).
\\
This procedure, that is defining a quantity $K_{\mu}$ which
fulfills (\ref{cond int1}), (\ref{cond int2}) and proposing a
scale invariant Lagrangian built of it, can be repeated in the
more complicated cases. Namely, Aratyn, Ferreira and Zimerman
\cite{aratyn} have introduced
\begin{equation}
K_{\mu}^{(2)}=(\partial_{\nu} u)^2 \partial_{\mu} u^* +\alpha
(\partial_{\nu} u \partial^{\nu} u^*) \partial_{\mu} u. \label{k2}
\end{equation}
Now, two cases are possible. If $\alpha \neq -1$ then the
integrability condition (\ref{cond int1}) leads to trivial
solutions. Otherwise, for $\alpha =-1$ the condition is always,
identically satisfied. Thus, any model based on $K^{(2)}_{\mu} $
with $\alpha =-1$ is integrable. It is in contradictory to the
Nicole-type models where equation (\ref{cond int1}) determines the
integrable (essentially restricted) submodel. The pertinent
Lagrangian reads
\begin{equation}
L=\frac{1}{(1+|u|^2)^3} (K_{\mu}^{(2)} \partial^{\mu}
u^*)^{\frac{3}{4}}. \label{aratyn}
\end{equation}
It has been checked that this model possesses infinitely many
toroidal solitons with arbitrary Hopf charge \cite{aratyn} (for
some generalization see \cite{ja AFZ}).
\\
In our work we would like to focus on the next possible form of
$K_{\mu}$. We assume it as follow
\begin{equation}
K_{\mu}^{(3)}= \alpha (\partial_{\nu} u \partial^{\nu} u^*)^2
\partial_{\mu} u +\beta (\partial_{\nu} u)^2 (\partial_{\nu} u
\partial^{\nu} u^*)
\partial_{\mu} u^* +\gamma (\partial_{\nu} u)^2(\partial_{\nu} u^*)^2
\partial_{\mu} u , \label{k3}
\end{equation}
where $\alpha, \beta, \gamma$ are real constants. The second
condition (\ref{cond int2}) is immediately fulfilled whereas the
first one (\ref{cond int1}) leads to an interesting formula.
Indeed, inserting (\ref{k3}) into (\ref{cond int1}) we get
\begin{equation}
(\partial_{\nu} u)^2 \left[ \gamma (\partial_{\nu} u)^2
(\partial_{\nu} u^*)^2 +(\alpha + \beta)(\partial_{\nu} u
\partial^{\nu} u^*)^2 \right] =0. \label{cond1 k3}
\end{equation}
It is not an identity unless $\alpha= - \beta $ and $\gamma =0$.
However, in this situation  $K^{(3)}_{\mu}$ is proportional to
$K^{(2)}_{\mu}$ and our problem can be reduced to the previously
discussed model. Thus, from now we assume that $\alpha \neq -
\beta $ and $\gamma \neq 0$ (for simplicity we assume that $\gamma
=1$). Then condition (\ref{cond1 k3}) can be rewritten in two
parts
\begin{equation}
(\partial u)^2=0 \label{cond2 eik}
\end{equation}
or
\begin{equation}
(\partial_{\nu} u)^2 (\partial_{\nu} u^*)^2 +(\alpha +
\beta)(\partial_{\nu} u \partial^{\nu} u^*)^2  =0 \label{cond2 k3}
\end{equation}
The first possibility is just the eikonal equation and does not
lead to new integrable submodels. However, the second equation
(\ref{cond2 k3}) (we called it generalized eikonal equation)
provides a new integrable condition for models based on
$K^{(3)}_{\mu}$.
\\
Let us now solve the generalized eikonal equation. Due to the fact
that we are mainly interested in obtaining of knotted
configurations with a nontrivial value of the Hopf index we
introduce the toroidal coordinates (\ref{tor_coord}) and assume
the following Ansatz
\begin{equation}
u(\eta, \xi, \phi) = f(\eta)e^{i(m\xi +n\phi)}. \label{anzatz}
\end{equation}
Then formula (\ref{cond2 k3}) gives
\begin{equation}
\left[ f'^2 - \left( m^2 +\frac{n^2}{\sinh^2 \eta} \right)f^2
\right]^2 +(\alpha+\beta) \left[ f'^2 + \left( m^2
+\frac{n^2}{\sinh^2 \eta} \right)f^2 \right]^2 =0, \label{cond3
k3}
\end{equation}
where prime denotes differentiation with respect to the $\eta$
variable and $f$ is a real shape function yet to be determined.
The last equation can be rewritten as
\begin{equation}
f'^4 - 2\frac{(\alpha +\beta -1)}{(\alpha +\beta+1)} f'^2f^2
\left( m^2 +\frac{n^2}{\sinh^2 \eta} \right) + f^4\left( m^2
+\frac{n^2}{\sinh^2 \eta} \right)^2=0 \label{cond4 k3}
\end{equation}
and possesses the following roots
\begin{equation}
f'^2=a^2(\alpha, \beta) \left( m^2 +\frac{n^2}{\sinh^2 \eta}
\right)f^2, \label{gen eik}
\end{equation}
where
\begin{equation}
a^2(\alpha, \beta)= \frac{(1-\alpha -\beta )}{(1+ \alpha +\beta)}
\pm \sqrt{\frac{(1-\alpha -\beta )^2}{(1+\alpha +\beta)^2}  -1}.
\label{pierwiastki}
\end{equation}
Equation (\ref{gen eik}) admits for a real solution only if $a^2
\geq 0$. Hence, we have a restriction for the parameters $\alpha$
and $\beta$ in $K^{(3)}_{\mu}$,
\begin{equation}
-1<\alpha + \beta \leq 0. \label{pierwiastki war}
\end{equation}
Finally we are able to solve (\ref{gen eik}). One can find that
\begin{equation}
f^{(\pm)}= A  \sinh^{\pm a |k|} \eta \frac{\left( |m| \cosh \eta +
\sqrt{k^2 +m^2\sinh^2 \eta } \right)^{\pm a|m|}}{\left( |k| \cosh
\eta + \sqrt{k^2 +m^2\sinh^2 \eta } \right)^{\pm a|k|}}.
\label{shape gen eik}
\end{equation}
In other words, we have obtained a solution of the generalized
eikonal equation
\begin{equation}
u(\eta, \xi, \phi) = A  \sinh^{\pm a |k|} \eta \frac{\left( |m|
\cosh \eta + \sqrt{k^2 +m^2\sinh^2 \eta } \right)^{\pm
a|m|}}{\left( |k| \cosh \eta + \sqrt{k^2 +m^2\sinh^2 \eta }
\right)^{\pm a|k|}}  e^{i(m\xi +n\phi)}. \label{sol gen eik}
\end{equation}
As the shape function smoothly interpolates from $0$ to $\infty$
(or from $\infty$ to $0$) this configuration carries the Hopf
topological charge $Q_H=\pm |mk|$. It is clearly visible that
difference between the standard eikonal knots and the generalized
solutions (\ref{sol gen eik}) emerges from the modified form of
the shape function. The original shape function $f_0$ in equation
(\ref{sol_eik}) is replaced by $f_0^{a}$ giving, depending on the
value of the model parameters $\alpha$ and $\beta$, squeezed or
stretched configurations.
\\
The generalized eikonal equation has similar symmetries as its
standard counterpart. As a consequence other solutions may be
obtained by a simple algebraic transformation
\begin{equation}
\tilde{u}=F (u), \label{sol gen eik ogolne}
\end{equation}
where $F$, identically as for the eikonal equation, is a
holomorphic function. Due to that, assuming that $F$ is a
polynomial, we are able to construct multi-knot configurations.
Namely,
$$
u(\eta, \xi, \phi) = $$
\begin{equation}
\sum_{j=1}^N A_j  \sinh^{\pm a |k_j|} \eta \frac{\left( |m_j|
\cosh \eta + \sqrt{k_j^2 +m_j^2\sinh^2 \eta } \right)^{\pm
a|m_j|}}{\left( |k_j| \cosh \eta + \sqrt{k_j^2 +m_j^2\sinh^2 \eta
} \right)^{\pm a|k_j|}}  e^{i(m_j\xi +n_j\phi)} +c_0
\label{knotted sol}
\end{equation}
where $A_j$ and $c_0$ are complex constants and the integer
parameters must obey the relation
\begin{equation}
\frac{m_j}{k_j}= \mbox{const.} \label{cond multi knot}
\end{equation}
Analogously as in the standard case, such a multi-knot solution
possesses the following topological charge (for configurations
with '-' sign) \cite{ja eikonal}
\begin{equation}
Q_H= - \mbox{max} \{ m_jk_j, \; \; j=1..N\}. \label{hopf index}
\end{equation}
Because of the fact that topological properties of the generalized
and standard eikonal knots do not differ drastically,  the
detailed geometrical analysis of these solutions can be found in
\cite{ja eikonal}.
\section{\bf{Integrable subsystems}}
Let us now introduce a new class of models based on the previously
defined quantity $K^{(3)}_{\mu}$ for which integrable subsystems
can be determined by the generalized eikonal equation. Moreover,
we will show that some members of the family possess generalized
eikonal knots as solutions of the dynamical equations of motion.
\\
The pertinent model reads
\begin{equation}
\mathcal{S}= \int d^4x \; G(|u|) \;  \left( K^{(3)}_{\mu}
\partial^{\mu} u^* \right)^{\frac{1}{2}}, \label{model}
\end{equation}
where $G$ is an arbitrary function of $|u|$ (sometimes called
'dielectric function').
\\
Then the field equation is
\begin{equation}
\partial_{\mu} \left[ G(|u|) \left( K^{(3)}_{\mu}
\partial^{\mu} u^* \right)^{-\frac{1}{2}} \mathcal{L}^{\mu} \right]-
\partial_{u^*} G(|u|) \left( K^{(3)}_{\mu}
\partial^{\mu} u^* \right)^{\frac{1}{2}}= 0, \label{eq mot1}
\end{equation}
where
\begin{equation}
\mathcal{L}_{\mu} = (1+\beta) (\partial u)^2 [2(\partial u
\partial u^*) \partial_{\mu} u^* +
(\partial u^*)^2 \partial_{\mu} u ] +3 \alpha (\partial u \partial
u^*)^2\partial_{\mu} u. \label{def L mu}
\end{equation}
One can notice that
\begin{equation}
\mathcal{L}_{\mu} \partial^{\mu} u^*=3 K_{\mu}^{(3)}\partial^{\mu}
u^* \label{propety1}
\end{equation}
and
\begin{equation}
\mathcal{L}_{\mu} \partial^{\mu} u=(\partial u)^2[
\left(2(1+\beta)+3 \alpha \right) (\partial u \partial u^*)^2 +
(1+\beta) (\partial u)^2 (\partial u^*)^2 ].\label{property2}
\end{equation}
In order to obtain an integrable submodel of (\ref{model}) we
impose the additional (integrability) condition \cite{aratyn}
\begin{equation}
\mathcal{L}_{\mu} \partial^{\mu} u=0, \label{property3}
\end{equation}
which leads to the generalized eikonal equation. Then, the
remaining equation of motion takes the following form
\begin{equation}
\partial_{\mu} \left[ G^{\frac{1}{3}}(|u|) \left( K^{(3)}_{\mu}
\partial^{\mu} u^* \right)^{-\frac{1}{2}} \mathcal{L}^{\mu}
\right]=0. \label{submodel1}
\end{equation}
In can be rewritten in the more compact form
\begin{equation}
\partial_{\mu} \mathcal{K}^{\mu}=0, \label{submodel2}
\end{equation}
where
\begin{equation}
\mathcal{K}^{\mu} = G^{\frac{1}{3}}(|u|) \left( K^{(3)}_{\mu}
\partial^{\mu} u^* \right)^{-\frac{1}{2}} \mathcal{L}^{\mu}.
\label{def K mu}
\end{equation}
These both equations: dynamical equation (\ref{submodel1}) and the
condition (\ref{property3}) define the integrable submodel. Here
integrability is understood as the existence of an infinite family
of the conserved current. In fact, using the results of
\cite{aratyn} one can show that such currents are given by the
expression
\begin{equation}
J_{\mu}= \mathcal{K}_{\mu} \frac{\partial H}{\partial
 u}-\mathcal{K}^*_{\mu} \frac{\partial H}{\partial
 u^*}, \label{current}
\end{equation}
where $H$ is an arbitrary function of $u$ and $u^*$.
\\
Now, we will prove that the existence of these currents can result
in the appearance of soliton solutions with a non-trivial value of
the Hopf index. At the beginning we specify a particular form of
the 'dielectric function' in the Lagrangian which allows us to
obtain toroidal solitons in the exact form. We take
\begin{equation}
G_{m,a}(|u|)= \left( \frac{ |u|^{\frac{1 -am}{am}
}}{1+|u|^{\frac{2}{am}}} \right)^3. \label{diel function}
\end{equation}
Taking into account Ansatz (\ref{anzatz}) we get
\begin{equation}
G_{m,a}(f)= \left( \frac{f^{\frac{1 -am}{am}
}}{1+f^{\frac{2}{am}}} \right)^3. \label{diel function f}
\end{equation}
Then the equation of motion (\ref{submodel1}) reads
\begin{equation}
\partial_{\eta} \left[ G^{1/3} I_0 c_+ f' \right] - G^{1/3}
I_0c_-f \left( m^2+\frac{k^2}{\sinh^2 \eta} \right) + G^{1/3}I_0
c_+f' \frac{\cosh \eta}{\sinh \eta}=0, \label{static eq1}
\end{equation}
where the following abbreviations have been made
\begin{equation}
I_0= \left[\left( (1+\beta) \omega_-^2 +\alpha \omega_+^2 \right)
\omega_+ \right]^{-\frac{1}{2}},\label{I0}
\end{equation}
\begin{equation}
c_{\pm}= \pm 2(1+\beta) \omega_-\omega_+ +(1+\beta) \omega_-^2 +
3\alpha \omega^2_+ \label{c}
\end{equation}
and
\begin{equation}
\omega_{\pm} = f'^2 \pm \left( m^2 +\frac{k^2}{\sinh^2 \eta}
\right) f^2. \label{omega}
\end{equation}
Of course, one should keep in mind that solutions of (\ref{static
eq1}) must obey the generalized eikonal equation as well. Hence
\begin{equation}
\omega_{\pm} = f'^2 \left( 1\pm \frac{1}{a^2} \right) \equiv
\lambda_{\pm} f'^2, \label{omega new}
\end{equation}
where
\begin{equation}
a^2=\frac{-3\alpha -\beta -1}{3(\alpha + \beta +1)} \pm
\sqrt{\left( \frac{3\alpha +\beta +1}{3(\alpha + \beta +1)}
\right)^2 -1}. \label{par a solution}
\end{equation}
Then
\begin{equation}
c_{\pm}= [\pm 2(1+\beta) \lambda_-\lambda_+ +(1+\beta) \lambda_-^2
+ 3\alpha \lambda^2_+] f'^{\; 4} \equiv \sigma_{\pm} f'^4 \label{c
new}
\end{equation}
and
\begin{equation}
I_0= \mbox{const.} \,  f'^{\, -3}. \label{I0 new}
\end{equation}
Inserting (\ref{c})-(\ref{I0 new}) into (\ref{static eq1}) one
obtains
\begin{equation}
\partial_{\eta} \ln ( G^{1/3} f'^2 ) -\frac{\sigma_-}{\sigma_+}
\frac{f}{f'} \left( m^2+\frac{k^2}{\sinh^2 \eta} \right) +
\partial_{\eta} \ln \sinh \eta =0, \label{static eq2}
\end{equation}
or equivalently
\begin{equation}
\partial_{\eta} \ln ( G^{1/3} f'^2 ) - \frac{\sigma_-}{a^2 \sigma_+}
\partial_{\eta} \ln f  +
\partial_{\eta} \ln \sinh \eta =0. \label{static eq3}
\end{equation}
Moreover, one can calculate that
\begin{equation}
\frac{\sigma_-}{\sigma_+} =a^2. \label{relation}
\end{equation}
Thus finally, equation (\ref{static eq3}) can be integrated and we
obtain
\begin{equation}
G^{1/3} f'^2f^{-1}= \frac{\mbox{const.}}{\sinh \eta} \label{eq F}
\end{equation}
It can be checked that this equation is solved by the following
function
\begin{equation}
f=\left( \frac{1}{\sinh \eta} \right)^{am}. \label{example}
\end{equation}
Since it also satisfies the generalized eikonal equation and
corresponds via (\ref{anzatz}) to the configuration with the
non-trivial topological charge $Q_H=-m^2$, we have proved that the
spectrum of soliton solutions of the integrable submodel is not
empty.
\\
In order to complete investigation of the generalized eikonal
hopfions we calculate their total energy. It can be performed in
the case of the hopfions for which the pertinent Lagrangian has
been established. In other words we do it for solutions
(\ref{example}) i.e. for knots with $m=k$.
\\
Then, the energy of the static configurations reads
\begin{equation}
E=\int d^3x G(|u|) (\vec{K}^{(3)} \cdot \vec{\nabla}
u^*)^{\frac{1}{2}}. \label{energy1}
\end{equation}
Taking into account that all solitons obey also the generalized
eikonal equation we obtain that
\begin{equation}
E=(2\pi)^2 \sqrt{\lambda_+ [ (1+\beta)\lambda_-^2+\alpha
\lambda_+]} \int_0^{\infty} d \eta \sinh \eta \left(
\frac{f^{\frac{1 -am}{am} }}{1+f^{\frac{2}{am}}} \right)^3 f'^3,
\label{energy2}
\end{equation}
or equivalently
\begin{equation}
E=(2\pi)^2 \sqrt{\lambda_+ [ (1+\beta)\lambda_-^2+\alpha
\lambda_+]} a^3 m^3 \int_0^{\infty} d \eta \frac{\cosh^3
\eta}{\sinh^2 \eta} \left( \frac{f^{\frac{1}{am}
}}{1+f^{\frac{2}{am}}} \right)^3. \label{energy3}
\end{equation}
Finally, after inserting (\ref{example}) into this formula we find
\begin{equation}
E=2\pi^2 \sqrt{\lambda_+ [ (1+\beta)\lambda_-^2+\alpha \lambda_+]}
a^3 m^3. \label{energy tot}
\end{equation}
As solutions (\ref{example}) possess the Hopf index $Q_H=-m^2$,
the energy depends on the topological charge in a rather
nontrivial manner i.e.
\begin{equation}
E = 2\pi^2 \sqrt{\lambda_+ [ (1+\beta)\lambda_-^2+\alpha
\lambda_+]} a^3 |Q_H|^{3/2}.
\end{equation}
One can notice that it resembles the energy-charge relation in the
modified Nicole models \cite{ja nicole}. It is nothing surprising
since the Nicole-type models can be achieved from introduced here
dynamical systems in the limit $a=1$.
\\
Such overlinear dependence found for many (generalized) eikonal
hopfions is rather puzzling and unexpected if we compare it with
the famous Va\-ku\-len\-ko-Ka\-pi\-tan\-sky energy-charge
inequality for the Fadd\-eev-Nie\-mi model \cite{vakulenko}, where
$E \geq c \, |Q_H|^{3/4}$. Moreover, it has been proved that this
sublinear dependence is valid also for the soliton solutions of
the Aratyn-Ferreira-Zimerman model \cite{aratyn}.
\\
Of course, it must to be emphasized that the overlinear behavior
is still only a conjecture. It is due to the fact that each
soliton (generalized eikonal hopfion) has been derived in a
different model. Thus, it is unknown whether for a fixed model
(i.e. fixed values of the parameters in (\ref{def L mu}) and
(\ref{diel function})) remained topological solutions obey this
relation. At this stage it would be hazardous to claim that all
introduced models must lead to overlinear dependence.
\section{\bf{Further models}}
The construction introduced and analyzed in the previous section
can be easily adopted to the more complicated (with a higher
number of derivatives) $K_{\mu}^{(i)}$.
\\
For example, let us focus on the next possibility and take
$$
K^{(4)}_{\mu}= \alpha (\partial u)^2 (\partial u^*)^2 (\partial u
\partial u^*) \partial_{\mu} u +\beta (\partial u
\partial u^*)^3 \partial_{\mu} u +  $$
\begin{equation}
\gamma (\partial u)^2 (\partial u
\partial u^*)^2 \partial_{\mu} u^* + \delta (\partial u)^4 (\partial u^*)^2\partial_{\mu}
u^*, \label{K4}
\end{equation}
where $\alpha, \beta, \gamma, \delta$ are real constants.
\\
The scale invariant action based on this quantity is given as
follows
\begin{equation}
S=\int d^4 x \tilde{G} (|u|) (K^{(4)}_{\mu} \partial^{\mu}
u^*)^{\frac{3}{8}}, \label{model 4}
\end{equation}
where $\tilde{G}$ is any function of $|u|$. The corresponding
integrable submodel can be found, in the analogous way as for
(\ref{model}), by imposing two conditions (\ref{cond int1}) and
(\ref{cond int2}). Since (\ref{cond int2}) is always fulfilled let
us turn to the first formula. One can show that, depending on the
parameters of the model, this equation leads to two cases. If
$\alpha = - \delta$ and $\beta = -\gamma$ then the condition is
identically satisfied. It is identically as in the
Aratyn-Ferreira-Zimerman model and no new solutions can be
obtained. Otherwise, we get an equation which is solved by
generalized eikonal knots. Thus, integrable subsystem of
(\ref{model 4}), though it exists, does not provide any new
hopfions.
\\
This feature appears to be quite general and one can observe it in
the more complicated examples.
\section{\bf{Conclusions}}
Let us shortly summarize the obtained results. A generalization of
the standard complex eikonal equation has been proposed. This
equation possesses various topologically non-trivial solutions. In
particular, knotted configurations carrying arbitrary value of the
Hopf charge have been explicitly derived. They appear to be
deformed (squeezed or stretched) standard eikonal knots. Moreover,
using the symmetry of the generalized eikonal equation, it is
possible to construct multi-knot solutions (linked knots).
\\
It has been also shown that this equation enables us to define a
new class of the integrable models, where integrability is
understood as the existence of an infinite family of the conserved
currents. Then, the generalized eikonal equation is just the
integrability condition. Additionally, we have proved that the
integrability may lead to the appearance of soliton solutions. In
the case of particular members of the analyzed family of the
models we have found that the spectrum of solutions is not empty
but consists of the generalized eikonal knots. Such Hopf solitons
i.e. hopfions have finite energy which depends on the topological
charge in a very nontrivial way.
\\
There are several directions in which one could continue the
present work. First of all, one should check whether, after fixing
the model parameters in (\ref{def L mu}), (\ref{diel function}),
the energy of other hopfions is also proportional to
$|Q_H|^{3/2}$. The validity of this conjecture (which could be
tested at least by some numerical methods) might be very important
in understanding of the interaction of Hopf solitons. Such a
overlinear dependence might suggest that, in these models,
hopfions with higher topological charges would decay into a
collection of the simplest solitons, each with the unit Hopf
index. It would be in contradictory to the standard clustering
phenomena observed in the Faddeev-Niemi model. Where, due to the
sublinear behavior, a separated multi-soliton configuration tends
to form a clustered, really knotted state \cite{ward}, \cite{lin}.
\\
Secondly, the generalized eikonal knots might be helpful in the
construction of new approximated, analytical solutions of the
Faddeev-Niemi hopfions (known only in the numeric form
\cite{helmut} \cite{battyde}, \cite{salo}), which could provide
better accuracy in the approximation of the energy.
\\
\vspace*{0.2cm}
\\
This work is partially supported by Foundation for Polish Science
FNP and ESF "COSLAB" programme.

\end{document}